\documentclass[twocolumn,showpacs,amsmath,amssymb,prl]{revtex4}

\usepackage{graphicx}

\begin{document}
\title{Intriguing Heat Conduction of a Polymer Chain}

\author{Jian-Sheng Wang} 
\affiliation{Singapore-MIT Alliance and Department of Computational Science,National University of Singapore, Singapore 117543, Republic of Singapore}

\author{Baowen Li}
\affiliation{
Department of Physics, National University of Singapore, Singapore 117542, Republic of Singapore}

\date{22 August 2003}

\begin{abstract}
We study heat conduction in a one-dimensional (1D) chain of particles
with longitudinal as well as transverse motions.  The particles are
connected by two-dimensional harmonic springs together with bending
angle interactions.  Using equilibrium and nonequilibrium molecular
dynamics, three types of thermal conducting behaviors are found: a
logarithmic divergence with system sizes for large transverse
coupling, $1/3$ power-law at intermediate coupling, and $2/5$
power-law at low temperatures and weak coupling.  The results are
consistent with a simple mode-coupling analysis of the same model.
The $1/3$ power-law divergence should be a generic feature for models
with transverse motions.
\end{abstract}

\maketitle

To understand the microscopic dynamical mechanism of heat conduction
is one of the long standing tasks in nonequilibrium statistical
mechanics. This problem has attracted increasing attention in recent
years [1--13]. The main effort has been focused on the necessary and
sufficient conditions of the Fourier law of heat conduction. With
strong numerical support, it is argued that chaos (or exponential
instability) is the necessary condition \cite{Casati}. However, recent
results show that even linear instability, such as that found in
generic polygonal billiards, is sufficient for a normal diffusion and
energy transport obeying the Fourier law \cite{mixing}.

In many systems studied so far, the heat conduction violates the
Fourier law, namely, the thermal conductivity diverges with system
size $N$ as $N^{\alpha}$, with $\alpha>0$. It has been proved that
for 1D system, the momentum conservation leads to an
anomalous heat conduction \cite{Prosen}.
However, its specific form of divergence
with system size is still of considerable controversy
\cite{1/3,2/5,lepri-mode-coupling,FPUExponent,Grassberger-Yang}. Based
on a renormalization group analysis for a 1D hydrodynamic fluid model,
it is argued that in a generic momentum conserving system, the thermal
conductivity should be $\kappa \propto
N^{1/3}$\cite{1/3}. Unfortunately, most existing numerical results do
not agree with this prediction because all 1D lattice models
considered so far have no transverse degree of freedom that is
required in the analysis. However, a mode-coupling theory analysis for
the 1D Fermi-Pasta-Ulam (FPU) model gives a divergent exponent $2/5$
\cite{2/5,lepri-mode-coupling}, which is supported by the numerics
from different groups \cite{FPUExponent}.

It seems that a universal exponent does not exist.  Most recently, it
is found that a divergent thermal conductivity is connected with a
superdiffusion \cite{LiWang}, $\kappa \propto N^{2-2/\beta}$, where
$\beta$ is the exponent of the diffusion ($\Delta x^2 \sim t^{\beta}$,
$0<\beta\le 2$). The value of $\beta$ changes from model to
model. This is justified by all billiard gas models studied.

On the other hand, the understanding of heat conduction mechanism will
allow us to control and manipulate heat current, and eventually to
design novel thermal devices with certain function
\cite{diode}.  To this end, more realistic physical models are
necessary.  Among many others, nanotubes and polymer chains are most
promising.  Recent molecular dynamics (MD) study of Carbon nanotubes
with realistic interaction potential suggested a divergent thermal
conductivity for narrow diameter tubes \cite{Nanotube}.  The strict 1D
models may not be applicable to nanotubes.  The added transverse
motion and the flexibility of the tube at long length scales 
will certainly scatter phonons, and thus should have a profound effect on
thermal transport.

In this Letter, we consider a polymer chain of $N$ point particles
with mass $m$ on a 1D lattice.  The lattice fixes the connectivity
topology such that only the neighboring particles interact.  The
Hamiltonian is given by
\begin{eqnarray}
H({\bf p}, {\bf r}) & = & \sum_{i} \frac{{\bf p}_i^2}{2m} +   
\frac{1}{2} K_r \sum_{i} \Bigl( | {\bf r}_{i+1}-{\bf r}_i | -a\Bigr)^2 
\nonumber
\\
&& +\, K_\phi \sum_{i} \cos \phi_i,
\end{eqnarray} 
where the position vector ${\bf r}=(x,y)$ and momentum vector ${\bf p}
=(p_x, p_y)$ are two-dimensional; $a$ is lattice constant.  If the
system is restricted to $y_i=0$, it is essentially a 1D gas with
harmonic interaction.  The coupling $K_r$ is the spring constant;
$K_\phi$ signifies bending or flexibility of the chain, while $\phi_i$
is the bond angle formed with two neighbor sites, $\cos
\phi_i = - {\bf n}_{i-1} \cdot {\bf n}_{i}$, and unit vector ${\bf n}_i = \Delta
{\bf r}_i /| \Delta {\bf r}_i|$, $\Delta {\bf r}_i = {\bf r}_{i+1} -
{\bf r}_{i}$.

We determine the heat current in a temperature gradient by
nonequilibrium MD.  The system is set up with fixed left most and
right most boundary.  The average distance between particles is set to
$a$, the zero-temperature equilibrium distance.  A group of four
particles at the two ends are subject to heat baths at temperature
$T_L$ and $T_H$, respectively. This is realized by Nos\'e-Hoover
thermostats.  The rest of the particles follow the equations of motion
using a velocity Verlet algorithm.  We use small time step sizes
$h=0.0005$ to $0.0010$.  Typical MD steps are $10^8$ to $10^{10}$.

\begin{figure}
\includegraphics[width=\columnwidth]{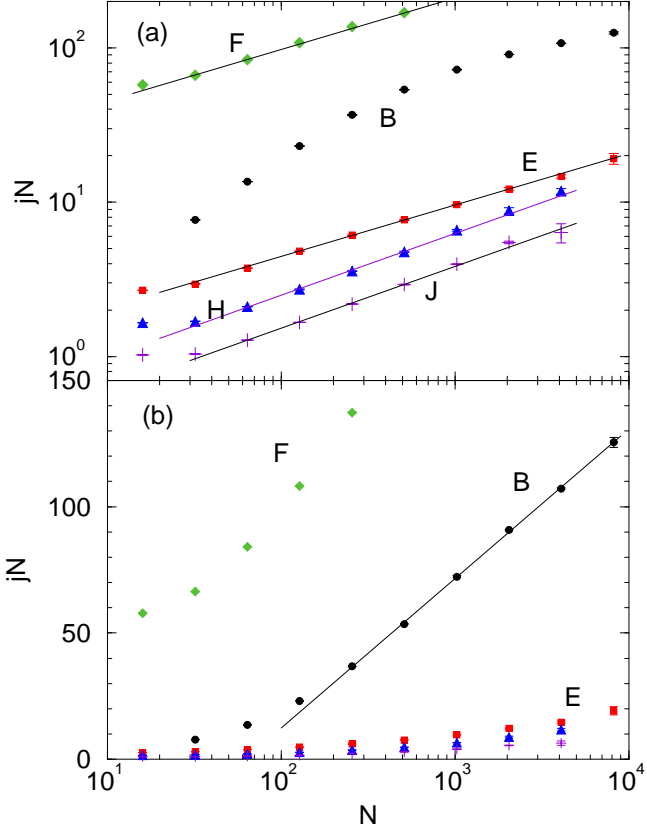}
\caption{\label{fig:jN} $jN$ vs $N$ on double-logarithmic (a)
and linear-log plot (b). The parameters  ($K_\phi, T_L, T_H$) of the model 
are,
set B: (1, 0.2, 0.4),
set E: (0.3, 0.3, 0.5),
set F: (1, 5, 7),
set H: (0, 0.3, 0.5),
set J: (0.05, 0.1, 0.2).  All of them have
$K_r = 1$, mass $m=1$, lattice spacing $a=2$.  The straight lines on F and E have
slope 1/3, while the slope on H and J is 2/5.
}
\end{figure}

We use the following expression for local heat current per particle:
\begin{eqnarray}
m\, {\bf j}_i &=&  -\, \Delta {\bf r}_i \bigl(( {\bf p}_i + 
{\bf p}_{i+1} ) \cdot {\bf G}(i) \bigr)
\nonumber\\ 
&& -\, \Delta {\bf r}_{i-1} \bigl( ( {\bf p}_i + {\bf p}_{i-1} ) 
\cdot {\bf G}(i-1) \bigr)
\nonumber\\
&& +\, \Delta {\bf r}_{i-1} \bigl( {\bf p}_i \cdot {\bf H}(i\!-\!2,i\!-\!1,i\!-\!1)\bigr)
\nonumber\\
&& +\, \Delta {\bf r}_{i} \bigl( {\bf p}_i \cdot {\bf H}(i\!+\!1,i\!+\!1,i) \bigr)
+ {\bf p}_i h_i, 
\end{eqnarray}
where ${\bf G}(i) = \frac{1}{4} K_r
\bigl(|\Delta {\bf r}_i | - a\bigr) {\bf n}_i$,
${\bf H}(i,j,k) = K_\phi( {\bf n}_i + {\bf n}_k \cos \phi_j) / |
\Delta {\bf r}_k |$, and the local energy per particle $h_i = 
\frac{1}{4} K_r \bigl( ( |\Delta {\bf r}_{i-1}|-a)^2 + (|\Delta {\bf r}_{i}|
-a)^2\bigr) + K_\phi \cos \phi_i + p_i^2/(2m)$.  This is derived from
${\bf J} = \sum d( {\bf r}_i h_i) /dt$, by regrouping some of the terms using
translational invariance.  It satisfies the
continuity equation in the long-wave limit.

We present our main numerical results in Fig.~\ref{fig:jN}.  The data
are obtained using 20 1GHz-Pentium PCs over six months of CPU times.
We plot the average heat current multiplied by $N$, $jN =(T_H - T_L)
\kappa$, in log-log scale (Fig.~\ref{fig:jN}(a)) and
linear-log scale (Fig.~\ref{fig:jN}(b)).  It is clearly shown that
three types of behaviors of the thermal conductivity $\kappa$ are
observed, the logarithmic divergence, $\log N$, power-law $\kappa
\propto N^\alpha$ with $\alpha = 1/3$, as well as $2/5$, depending on
the model parameters.  Log-log plot shows linear behavior for data set
E, F, H, and J.  At the parameters of set E, excellent power-law
dependence is found, with an exponent of $\alpha = 0.334 \pm 0.003$
(using an error weighted least-squares fit for $N\ge 128$).  Set F is
also in good agreement with a slope of 1/3.  On the other hand, for
set H and J, we have exponent $\alpha$ consistent with 0.4.  Set B is
consistent with logarithmic divergence, $\kappa \propto \log N$ (see
Fig.~\ref{fig:jN}(b)).  The model has two key parameters, the
temperature $T$, and the transverse coupling $K_\phi$.  We should
mention that wide range of parameters is scanned, and surprisingly,
only the three scalings are found so far in this model.

To understand the simulation results, we consider a simple
mode-coupling theory for the present model.  The equations of motion
in terms of normal-mode coordinates, $Q_k^\parallel =
\sqrt{\frac{m}{N}} \sum_{j=0}^{N-1} (x_j-ja) e^{i2\pi jk/N}$,
$Q_k^\perp = \sqrt{\frac{m}{N}} \sum_{j=0}^{N-1} y_j e^{i2\pi jk/N}$,
for small oscillation near zero-temperature equilibrium position,
keeping only leading nonlinearity, are
\begin{subequations}
\begin{eqnarray}
{d^2Q_k^\parallel \over dt^2} &=& - (\omega_k^\parallel)^2 Q_k^\parallel 
 + \!\!\!\sum_{k'+k''=k}\!\!\! c_{k',k''}^\parallel Q_{k'}^\perp Q_{k''}^\perp,
\\
{\ d^2Q_k^\perp \over dt^2} &=& - (\omega_k^\perp)^2 Q_k^\perp 
 + \!\!\!\sum_{k'+k''=k}\!\!\! c_{k',k''}^\perp Q_{k'}^\parallel Q_{k''}^\perp,
\end{eqnarray}
\end{subequations}
where the bare dispersion relations are given by $\omega_k^\parallel =
2 \sqrt{\frac{K_r}{m}} | \sin \frac{\pi k}{N}|$, and $\omega_k^\perp =
4 \sqrt{\frac{K_\phi}{a^2m}} \sin^2\frac{\pi k}{N}$.  The expressions
for $c_k^{\parallel,\perp}$ are complicated, but can be simplified in
the long-wave limit, as $c_{k,k'}^\perp = 2 c_{k,k'}^\parallel \propto
kk'(k+k')$.  Instead of the integer $k$, we can also index the mode by
its corresponding lattice momentum $p = 2\pi k/(aN)$.

A central quantity in the mode-coupling theory is the normalized
correlation function, $g_p(t) = \langle Q_p(t) Q_{p}^*(0) \rangle/
\langle |Q_p(0)|^2 \rangle$.  The Fourier-Laplace transform of the
correlation function, $g[z] = \int_0^\infty g(t) e^{-izt}\, dt$, is
given by
\cite{Kubo,Lepri-review},
\begin{equation}
  g^{\parallel, \perp}_p[z] = 
{ -iz - p^2 \nu^{\parallel,\perp}[z] 
\over z^2 - c_{\parallel,\perp}^2 p^2 - i z\, p^2 \nu^{\parallel,\perp}[z] }. 
\label{Geq}
\end{equation}
The constants $c_\parallel$ and $c_\perp$ are effective or
renormalized sound velocities for the longitudinal and transverse
modes.  They are defined, e.g., by $(c_\parallel p)^2 \langle
|Q_p^\parallel|^2 \rangle = k_BT$, as $p \to 0$.  The damping
functions (memory kernel) are given by the coupled equations,
\begin{subequations}
\label{veq}
\begin{eqnarray}
 \nu^\parallel(t) &=& { K_\parallel \over 2\pi} 
\int_{-\frac{\pi}{a}}^{\frac{\pi}{a}}\!\!\! dp\: \bigl(g_p^\perp(t)\bigr)^2 ,
\label{veqa}
\\
 \nu^\perp(t) &=& { K_\perp \over 2\pi} \int_{-\frac{\pi}{a}}^{\frac{\pi}{a}} 
\!\!\!dp\: g_p^\parallel(t)\, g_p^\perp(t) .
\label{veqb}
\end{eqnarray}
\end{subequations}
Eqs. (\ref{Geq}) and (\ref{veq}) form a closed system of nonlinear
integral equations.  This is a straightforward generalization of the
strict 1D result \cite{lepri-mode-coupling}.  The above equations are
derived under a number of simplification assumptions, such as
long-wave approximation, mean-field type product approximation for the
correlation functions, replacing random-force correlation with true
force correlation. Some of them can be removed but more complicated
equations will result.

\begin{figure}
\includegraphics[width=\columnwidth]{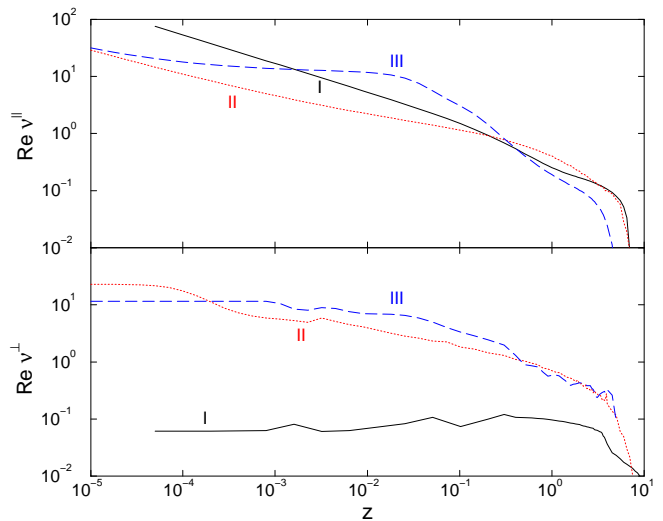}
\caption{\label{fig:mode} Real part of $\nu^\parallel[z]$ and 
$\nu^\perp[z]$ vs $z$ for parameters $a=1$, $K_\parallel=1$, and
($K_\perp$, $c_\parallel$, $c_\perp$) I: (0.3, 2, 1), II: (1.8, 1, 1),
III: (2, 1, 0.5).}
\end{figure}

In Fourier space, for large $z$ the solution is found from integration
by part, as $\nu^{\parallel,\perp}[z] = K_{\parallel,\perp}/(iza) +
O(z^{-3})$.  The long-wave asymptotic decay of each mode is controlled
by the small $z$ behavior of the function $\nu^{\parallel, \perp}[z]$.
We define $\delta_\parallel$ and $\delta_\perp$ by
$\nu^{\parallel,\perp}[z] \propto z^{-\delta_{\parallel,\perp}}$.  The
dispersion relation is then given by the location of the poles in the
correlation function $g[z]$.  The imaginary part of the frequency
gives damping, by $\gamma_p \propto p^2
\nu[z]\big|_{z \to cp} \propto p^{2-\delta}$.  We note that three types of
behaviors can be derived from the above set of equations.  If
$K_\parallel \approx K_\perp$ and $c_\parallel \approx c_\perp$, the
two equations reduce to that of strict 1D model, we thus
expect the result of Lepri \cite{lepri-mode-coupling}, 
i.e., $\delta_\parallel = \delta_\perp =
1/3$.  One the other hand, it can be shown rigorously that in the
limit of small $K_\perp$ and small $c_\perp$, we have $\delta_\perp
=0$ and $\delta_\parallel = 1/2$.  Formally, when $a \to 0$, the
equation possesses the scaling solution of the form $\nu[\lambda z] =
\lambda^{-1} \nu[z]$; this implies $\nu[z] \propto 1/z$.  
These analytic results are supported by numerical solutions of the
coupled equations, shown in Fig.~\ref{fig:mode}.  They are solved by a
brute force numerical integration in Fourier space.
Details of the mode-coupling calculation will be presented elsewhere.

In Fig.~\ref{fig:mode} at parameter set I, we observe very good
asymptotic behavior of $\nu^\parallel[z] \propto z^{-1/2}$ and
$\nu^\perp[z] \propto {\rm const}$. This corresponds to the behavior
of MD results for data set E and F in Fig.~\ref{fig:jN}.  When
$c_\parallel = c_\perp$ but $K_\parallel \neq K_\perp$ (set II), we
see crossover from $\delta_\parallel = 1/3$ to $1/2$.  The curve III
may be related to the logarithmic divergence.  We note that a
meaningful, direct mapping from the simulation parameters to
mode-coupling parameters is not possible, due to qualitative nature of
the theory.

\begin{figure}
\includegraphics[width=\columnwidth]{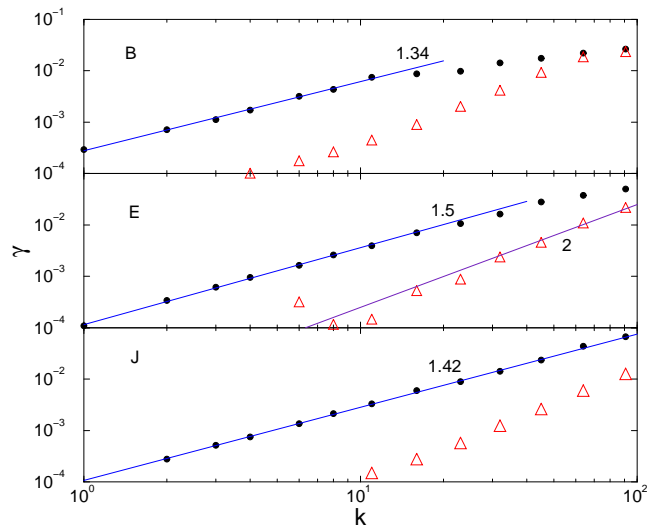}
\caption{\label{fig:gamma} The decay rate $\gamma^\parallel$ (dots) and 
$\gamma^\perp$ (triangles) vs $k$ for the parameters set B, E, and J at
equilibrium temperature $T=(T_L+T_H)/2$.  The number on the line
indicates the slope of the straight line.  The system size is
$N=1024$.}
\end{figure}

The prediction of $\delta_\parallel = 1/2$ and $\delta_\perp = 0$ is
checked against an equilibrium MD simulation in a microcanonical
ensemble with periodic boundary condition.  We compute the normal-mode
correlation $\langle Q_p(t) Q_p^*(0) \rangle$ for each mode specified
by the lattice momentum $p = 2\pi k/(aN)$.  The functions are
oscillatory with an exponential decay, $\cos(\omega t) e^{-\gamma_p
t}$.  The decay constants are obtained by fitting the maximum amplitude
as a function of time.  The results are presented in
Fig.~\ref{fig:gamma}.  Comparing with results from smaller and larger
system sizes, effect of finite sizes appears rather small at $N=1024$.
Excellent agreement with mode-coupling theory ($\gamma \propto
k^{2-\delta}$) is obtained for data set E.  However, for data set B
and J, the slopes are not consistent with either logarithmic
divergence for $\kappa$ or 2/5 law.  This may be interpreted as that
we are still not in the asymptotic regime.

To connect the result of damping of the modes with thermal
conductivity, it is noted \cite{2/5,perverzev} that each mode
contributes to the thermal transport independently.  Under the
linear-response theory, the Green-Kubo formula relates the
current-current correlation to the thermal conductivity by
\begin{equation}
 \kappa = \frac{1}{k_B T^2 a N} \int_0^\infty \langle J(t) J(0) \rangle dt.
\end{equation}
The decay rates for $J$ are assumed to be the same as that for $Q$,
thus $\langle J(t)J(0)\rangle \propto \sum_p \exp(-\gamma_p^\parallel
t)$.  The amplitude of the exponential decay is approximately
independent of $p$.  Converting the summation to integral, we have $
\langle J(t) J(0)\rangle \propto t^{-1/(2-\delta_\parallel)}$.  The
thermal conductivity on a finite lattice is obtained by integrating
over $t$ to a time of $O(N)$, Thus $\kappa_N \propto
N^{1-1/(2-\delta_\parallel)} = N^\alpha$.  When $\delta_\parallel =
1/2$, we have $\alpha = 1/3$, and when $\delta_\parallel = 1/3$, the
exponent $\alpha = 2/5$.

\begin{figure}[t]
\includegraphics[width=\columnwidth]{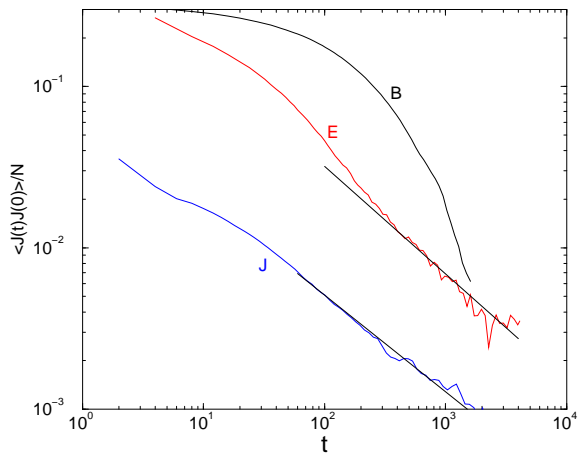}
\caption{\label{fig:GK} The Green-Kubo integrand,
$\langle J(t) J(0) \rangle/N$, vs time $t$.  The parameters are the
same as that in Fig.~\protect\ref{fig:gamma}.  The straight lines have
slope $-2/3$ (on E) and $-3/5$ (on J).}
\end{figure}

The current-current correlation functions are presented in
Fig.~\ref{fig:GK} for the parameters corresponding to data set B, E,
and J in Fig.~\ref{fig:jN}.  For data set J, a power-law dependence is in excellent
agreement with the theoretical expectation $t^{\alpha-1}$ with
$\alpha=2/5$.  For set E, the curve is a bit steeper than expected.
This may be due to finite sizes. For set B, where logarithmic divergence
is observed, we do not observe good power-law behavior in the
correlation.

We need to clarify the relationship between the three types of
observed behaviors in the nonequilibrium MD results.  From
mode-coupling point of view, the 1/3-law is generic and robust, while
$\alpha=2/5$ should be eventually crossover to 1/3 at long length
scales.
However, such a crossover is not observed in MD data.  The crossover
effect can be argued for a more general setting.  More general
mode-coupling equations for a generic interaction potential consistent
with the symmetry would have an additional term of ${ K_3 \over 2\pi}
\int_{-\frac{\pi}{a}}^{\frac{\pi}{a}}
dp\bigl(g_p^\parallel(t)\bigr)^2$ for Eq.~(\ref{veqa});
Eq.~(\ref{veqb}) remains the same.  
Such a term can appear either from cubic or quartic nonlinearity 
in potential.  Contribution from this extra term
decays in time $t$ faster than the perpendicular component
contribution.  Thus, the asymptotic result of $\delta_\parallel = 1/2$
remains true.  The same should be also true even if a chain is allowed
to move in three dimensions.  If the parameter $K_3$ is sufficiently
large, we may see exponent close to $0.4$.  The logarithmic divergence
is a bit difficult to interpret.  It may require a more refined theory
than the present naive mode-coupling theory.

In summary, we have observed three different scalings in a 1D polymer
chain.  When the transverse motion couples with the longitudinal
motion, the thermal conductivity diverges with system size with a 1/3
power-law.  This has been demonstrated with a very high precision
numerical result and explained in terms of a mode-coupling theory.  In
the weak coupling regime, a 2/5 power-law is observed which is
consistent with the results observed in the FPU model. In the case of
strong transverse coupling, a logarithmic divergent law is observed.

JSW is supported in part by Singapore-MIT Alliance and by the Academic
Research Fund (ARF) of National University of Singapore (NUS). BL is
supported in part by ARF of NUS.  We thank Dr.~Weizhu Bao for
discussion.


\begin{thebibliography}{01}

\bibitem{Casati} G. Casati \textsl{et al.},
Phys. Rev. Lett. \textbf{52}, 1861 (1984); D. Alonso \textsl{et al.},
\textsl{ibid.} \textbf{82}, 1859 (1999).

\bibitem{mixing} B. Li \textsl{et al.}, Phys. Rev. Lett. \textbf{88}, 223901
(2002); B. Li \textsl{et al.}, Phys. Rev. E \textbf{67}, 021204
(2003); B. Li \textsl{et al.}, cond-mat/0307692; D. Alonso \textsl{et
al.}, Phys. Rev. E \textbf{66}, 066131 (2002).

\bibitem{Prosen} T. Prosen and D. K. Campbell, Phys. Rev. Lett.  \textbf{84}, 2857 (2000).

\bibitem{1/3} O. Narayan and S. Ramaswamy, Phys. Rev. Lett.  \textbf{89}, 200601 (2002).

\bibitem{2/5} S. Lepri \textsl{et al.} Europhys. Lett. \textbf{43}, 271 (1998).

\bibitem{lepri-mode-coupling} S. Lepri, Phys. Rev. E \textbf{58}, 
7165 (1998).

\bibitem{FPUExponent} H. Kaburaki and M. Machida, Phys. Lett. A\textbf{181}, 85 (1993);
S. Lepri \textsl{et al}, Phys. Rev. Lett. \textbf{78}, 1896 (1997);
B. Hu \textsl{et al.}, Phys. Rev. E \textbf{61}, 3828 (2000).

\bibitem{Grassberger-Yang} P. Grassberger et al, Phys. Rev. Lett. \textbf{89} 180601 (2002); G. Casati and T. Prosen, Phys. Rev. E \textbf{67}, 015203 (R) (2003). 

\bibitem{LiWang} B. Li and J. Wang, Phys. Rev. Lett. \textbf{91}, 044301 (2003).

\bibitem{diode} M. Terraneo \textsl{et al.} Phys. Rev. Lett \textbf{88}, 094302 (2002).  

\bibitem{Nanotube} S. Maruyama, Physica B \textbf{323}, 193 (2002).

\bibitem{Lebowitz} F. Bonetto \textsl{et al.}, 
in ``Mathematical Physics 2000,'' A. Fokas \textsl{et al.} (eds)
(Imperial College Press, London, 2000) (pp. 128-150).

\bibitem{Lepri-review} S. Lepri \textsl{et al.}, Phys. Rep. 
\textbf{377}, 1 (2003). 

\bibitem{Kubo} R. Kubo \textsl{et al}, \textsl{Statistical
Physics II}, pp.97-108, 2nd ed. Springer, Berlin (1992).

\bibitem{perverzev} A. Perverzev, cond-mat/0306747.

\end{thebibliography}
\end{document}